\documentclass[onecolumn,letterpaper,11pt,draftclsnofoot]{IEEEtran}
\usepackage{times,amsmath,epsfig}
\usepackage{cite}
\usepackage{graphicx}
\usepackage{psfrag}
\usepackage{subfigure}
\usepackage{url}
\usepackage{stfloats}
\usepackage{amsmath}
\usepackage{amssymb}
\usepackage{array}
\usepackage{array}

\newtheorem{theorem}{Theorem}

\newtheorem{lemma}{Lemma}

\newtheorem{definition}{Definition}

\ifCLASSINFOpdf
\else
\fi
\begin{document}
\title{\huge{Non-cooperative Feedback Rate Control Game for Channel State Information in Wireless Networks}}

%% author names and affiliations
%% use a multiple column layout for up to three different
%% affiliations
%\author{\IEEEauthorblockN{Lingyang Song}
%\IEEEauthorblockA{School of Electrical Engineering and Computer Science\\
%Peking University, Beijing, China\\
%Email: lingyang.song@pku.edu.cn}
%\and
%\IEEEauthorblockN{Zhu Han}
%\IEEEauthorblockA{Electrical and Computer Engineering Department\\
%                  University of Houston, Houston, US\\
%Email: zhan2@mail.uh.edu}
%}

\author{
\IEEEauthorblockN{Lingyang~Song\IEEEauthorrefmark{1}, Zhu~Han\IEEEauthorrefmark{2}, Zhongshan~Zhang\IEEEauthorrefmark{3}, and Bingli~Jiao\IEEEauthorrefmark{1}\vspace{3mm}\\}
\IEEEauthorblockA{\IEEEauthorrefmark{1}School of Electronics
Engineering and Computer Science, \vspace{-3mm}\\
Peking University, Beijing, China.
\\ \IEEEauthorrefmark{2}Electrical and Computer Engineering Department,\vspace{-3mm} \\ University of Houston, Houston, USA.
\\ \IEEEauthorrefmark{3}Department of Wireless Communications,\vspace{-3mm} \\ NEC Laboratories, Beijing, China.
\\
%Email: lingyang.song@pku.edu.cn; \ zhan2@mail.uh.edu
}\thanks{This paper is partially supported by US NSF CNS-0910461, CNS-0905556, CNS-0953377, and ECCS-1028782.}}


%Telephone: (800) 555--1212, Fax: (888) 555--1212}


\maketitle

\thispagestyle{empty}
\begin{abstract}
%\boldmath
It has been well recognized that channel state information~(CSI) feedback is of great importance for dowlink transmissions of closed-loop wireless networks. However, the existing work typically researched the CSI feedback problem for each individual mobile station~(MS), and thus, cannot efficiently model the interactions among self-interested mobile users in the network level. To this end, in this paper, we propose an alternative approach to investigate the CSI feedback rate control problem in the analytical setting of a game theoretic framework, in which a multiple-antenna base station~(BS) communicates with a number of co-channel MSs through linear precoder. Specifically, we first present a non-cooperative feedback-rate control game~(NFC), in which each MS selects the feedback rate to maximize its performance in a distributed way. To improve efficiency from a social optimum point of view, we then introduce pricing, called the non-cooperative feedback-rate control game with price~(NFCP). The game utility is defined as the performance gain by CSI feedback minus the price as a linear function of the CSI feedback rate. The existence of the Nash equilibrium of such games is investigated, and two types of feedback protocols (FDMA and CSMA) are studied. Simulation results show that by adjusting the pricing factor, the distributed NFCP game results in close optimal performance compared with that of the centralized scheme.
\end{abstract}


% \fi
\IEEEpeerreviewmaketitle

\newpage  \setcounter{page}{1}\setlength{\baselineskip}{25pt}

%%%%%%%%%%%%%%%%%%%%%%%%%%%%%%%%%%%%%%%%%%%%%%%%%%%%%%%%%%%%%%%%%%%%%%%%%%%%%%%%%%%%%%%%%%%%%%%%%%%
\section{Introduction}
%%%%%%%%%%%%%%%%%%%%%%%%%%%%%%%%%%%%%%%%%%%%%%%%%%%%%%%%%%%%%%%%%%%%%%%%%%%%%%%%%%%%%%%%%%%%%%%%%%%
The increasing demands for fast and reliable wireless communications have spurred development of multiple-antenna systems in order to efficiently harvest the capacity gains~\cite{Telatar99,Tarokh98}. Recent information-theoretic research indicates that a feedback channel can be further employed to furnish channel state information (CSI) to the transmitter side, which may affect closed-loop capacity gains~\cite{Ref:David-ovwerview}. With some form of knowledge of the wireless channel conditions, the transmitter can adapt to the propagation conditions by the use of a variety of channel adaptive techniques~\cite{Paulraj-STWC}. Specifically, in a multiple mobile station (MS) scenario, with the knowledge of the channel to nearby co-channel MSs, it is possible to actively suppress the signal to the interfered users and meanwhile maximize the effective signal power~\cite{Ref:David-value}. In this case, the base station~(BS) can obtain the required channel coefficients through a feedback channel from the MS. Then, mechanisms such as multiple-antenna precoding can be utilized to mitigate the effects of co-channel interference and exploit spatial dimensions to increase the capacity of wireless networks~\cite{Rashid-Farrokhi1998}.

Since CSI is essential for closed-loop wireless communication systems, the techniques on how to effectively feedback CSI from the transmitter to the receiver has been intensively studied~\cite{Paulraj-STWC,Ref:David-ovwerview}. As perfect feedback of CSI is typically unavailable due to complexity or practicality constraints, the infinite feedback of CSI is hard to realize in practice. Therefore, it is important to investigate how to control the amount of feedback signalling overhead according to the individual requirements in order to achieve good quality of service~(QoS). As a result, CSI feedback rate control problem has attracted lots of attention in recent years~\cite{Ref:David-value,Ref:T.Eriksson-summary-of-compression}. In~\cite{Bhashyam2002}, the quantized feedback approach for power-control is designed to minimize an upper bound of multiple-input-single-output (MISO) systems. Recently, two specific forms of partial feedback, namely, channel mean feedback~\cite{Zhou002} and channel covariance feedback~\cite{Simon2003}, have been investigated for slow-varying and rapidly varying MIMO channels, respectively.

The existing work typically treated each MS independently, and researched the multi-MS CSI feedback problem in physical layer, e.g., from either communication or information theory point of view. This cannot efficiently model the interactions among self-interested mobile users in wireless systems~\cite{Ref:David-value,Ref:David-ovwerview}. If the feedback channel is limited, there exists conflicts in effective CSI feedback rates between each MS. If one MS transmits too much CSI, it will result in the reduction of the rest MSs' CSI feedback amounts, and thus, degrade the others' performance. Hence, it will be desirable to sort out the competition problem by finding a balance in this multi-MS feedback scenario, and meanwhile achieve better QoS.

Game theory \cite{Fudenberg1993,Maskery2009} offers a set of mathematical tools to study the complex interactions among interdependent rational players and predict their choices of strategies~\cite{Mas-Colell1995,Attar2008,Attar2009}. In this paper, an alternative approach to the feedback rate control problem in wireless systems based on an economic model is proposed. In this model, each MS's preference is represented by a utility function, which quantifies the level of satisfaction a user gets from using the system resources~\cite{Shah1998}. Each player in the game maximizes a utility function in a distributed fashion. The game settles at a Nash equilibrium if one exists. Since users act selfishly, the equilibrium point is not necessarily the best operating point from a social point of view~\cite{Saraydar2002}. To achieve a more socially desirable result, a powerful tool by pricing the system resources can be introduced, which is able to guide user behavior toward a more efficient operating point~\cite{Shah1998,Saraydar2002}.

To the best of our knowledge, the game-theoretic methods are first applied to study CSI feedback rate control under this economical model. Specifically, we investigate the scenario in a single-cell wireless data network, where a multiple-antenna BS communicates with a number of co-channel users through a minimum mean square error~(MMSE) precoder and each user tries to maximize its own utility. Two types of feedback protocols, FDMA and CSMA, are investigated. For ease of understanding, we first present a noncooperative feedback rate control game~(NFC), which optimizes individual utility in a distributed fashion. While the resulting noncooperative feedback rate control game has a Nash equilibrium, it is inefficient from a social point of view. Therefore, we further introduce pricing to create cooperation between each MS in order to improve efficiency, called the noncooperative feedback-rate control game with price~(NFCP). The price function is a linear function of the CSI feedback rate that also allows a distributed implementation by broadcasting the price per bandwidth from the BS to all the MSs. It shows that there exists an equilibrium in the proposed NFCP. Simulation results indicate that by adjusting price, NFCP provides better overall utility than NFC. In addition, the distributed NFCP game approach achieves near optimal performance compared to the centralized scheme, and thus, improves the overall throughput of wireless data networks.

The rest of the paper is organized as follows: In Section~II, we introduce some preliminaries, including the system model, the multiple access protocols, and CSI feedback rate model. In Section~III, we describe the MMSE precoder, and some properties by using the CSI feedback rate model. The proposed NFC and NFCP algorithms are described in~IV. Simulation results are provided in Section~V. In Section~VI, we draw the main conclusions. Some derivations are given in the appendixes.

\emph{\textbf{Notation}}: Boldface lower-case letters denote vectors, $(\cdot)^{*}$, $(\cdot)^{T}$ and $(\cdot)^{H}$ represent conjugate, transpose, and conjugate transpose, respectively, $\|\mathbf{x}\|^2=\mathbf{x}^H\mathbf{x}$, and $\mathrm{Var}[x]$ represents its variance.

%%%%%%%%%%%%%%%%%%%%%%%%%%%%%%%%%%%%%%%%%%%%%%%%%%%%%%%%%%%%%%%%%%%
\section{Preliminaries}
%%%%%%%%%%%%%%%%%%%%%%%%%%%%%%%%%%%%%%%%%%%%%%%%%%%%%%%%%%%%%%%%%%%
%%%%%%%%%%%%%%%%%%%%%%%%%%%%%%%%%%%%%%%%%%%%%%%%%%%%%%%%%%%%%%%%%%%

In this section, we first give the system model. Then two types of feedback channels are discussed. Finally, the CSI feedback rate model is illustrated.

\subsection{System Model}
\begin{figure}[]
\centering
\includegraphics[height=2.9in,width=3.2in]{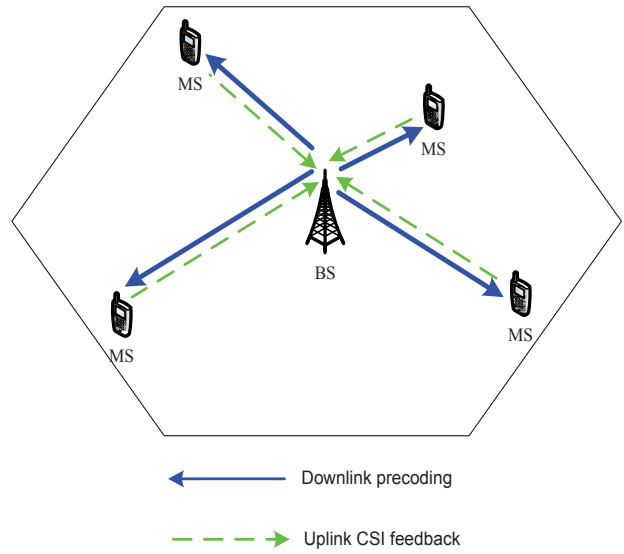}
\caption{System model: One BS is serving a number of MSs by precoding based on the CSI feedback.} \label{Fig:scenario}
\end{figure}

%%%%%%%%%%%%%%%%%%%%%%%%%%%%%%%%%%%%%%%%%%%%%%%%%%%%%%%%%%%%%%%%%%%
In this paper, we consider a system in which a number of co-channel MSs are served by one BS. The architecture is depicted in Fig.\ref{Fig:scenario}. The BS is assumed to know the linear processing performed by the MSs, which can acquire the required CSI through a feedback channel from the MSs. Using multiple antennas at the BS of a cellular system, transmit precoding can be performed for simultaneous transmission to several co-channel mobile users. The precoder is designed assuming a stationary scenario in which the fast (Rayleigh) fading is described by its second order properties. We also assume narrow-band signals without any time dispersion, i.e., the channel fading is frequency flat. For simplicity, we assume every MS is equipped with a single receive antenna. We assume that the system works in a FDD model, where the BS has $N_t$ transmit antennas serving $N_s$ MSs simultaneously in the same frequency band, while each MS feeds back the CSI through different channels in order to better protect the control information by avoiding collisions.

For the $k$-th MS, the input signal, $x_k$, is first precoded by complex weights $\mathbf{w}_k\in\mathbb{C}^{N_t\times{1}}$ before transmitted from the $N_t$ antennas at the BS. The corresponding output can be written as
%%--
\begin{equation}
    \textbf{s}_k=\textbf{w}_kx_k,
    \label{Eq:Sopj}
\end{equation}
%%--
where $\textbf{s}_k\in\mathbb{C}^{N_t\times{1}}$.

The received signal at the $k$-th MS can be then expressed as
%%--
\begin{align}
    y_k
    &=\mathbf{h}_{k}^T\sum_{i=1}^{N_s}\mathbf{s}_i+n_k
    \nonumber \\
    &=\mathbf{h}_{k}^T\sum_{i=1}^{N_s}\mathbf{w}_i^Hx_i+n_k
    \nonumber \\
    &=\mathbf{h}_{k}^T\mathbf{w}_kx_i+\mathbf{h}_{k}^T\sum_{i=1,i{\neq}k}^{N_s}\mathbf{w}_ix_i+n_k,
    \label{Eq:Sui}
\end{align}
%%--
where $\mathbf{h}_{k}=[h_{1,k},\ldots,h_{N_t,k}]^T\in\mathbb{C}^{N_t{\times}1}$ represents the channel coefficients from the BS to the $k$-th MS with zero mean and unit variance, $n_k$ is the AWGN noise $\mathcal{CN}(0,N_0)$, $\mathbf{h}_{k}^T\mathbf{w}_kx_k$ is the desired signal, and $\mathbf{h}_{k}^T\sum_{i=1,i{\neq}k}^{N_s}\mathbf{w}_ix_i$ can be treated as the interference. Note that the model can easily be extended to frequency selective channels, taking both co-channel interference and inter-symbol interference into account \cite{Rashid-Farrokhi1998}.

%%%%%%%%%%%%%%%%%%%%%%%%%%%%%%%%%%%%%%%%%%%%%%%%%%%%%%%%%%%%%%%%%%%
\subsection{Two Multiple Access Models}
%%%%%%%%%%%%%%%%%%%%%%%%%%%%%%%%%%%%%%%%%%%%%%%%%%%%%%%%%%%%%%%%%%%
In this subsection, two multiple access protocols for the uplink are presented. For simplicity, in the downlink, the BS simultaneously serves multiple co-channel mobile users by performing precoding, while in the uplink two standard multiple access protocols are examined for CSI feedback. We assume that the total system bandwidth is $B$, and the bandwidths for downlink and uplink transmissions are $B_{DL}$ and $B_{UL}$, respectively. Then, we have
\begin{equation}
    B = B_{DL} + B_{UL}.
    \label{Eq:bwconstaint}
\end{equation}
%%--
%%%%%%%%%%%%%%%%%%%%%%%%%%%%%%%%%%%%%%%%%%%%%%%%%%%%%%%%%%%%%%%%%%%
\subsubsection{Frequency Division Multiple Access}
%%%%%%%%%%%%%%%%%%%%%%%%%%%%%%%%%%%%%%%%%%%%%%%%%%%%%%%%%%%%%%%%%%%
The BS serves MSs simultaneously in the same frequency band, while each MS feeds back the CSI through orthogonal channels, i.e., frequency division multiple access~(FDMA), in order to better protect the control information by avoiding interference.

Recalling (\ref{Eq:rat}), the uplink bandwidth can be then calculated as
%%--
\begin{equation}
    B_{UL}=\beta\sum_{k=1}^{N_s}r_k,
    \label{Eq:Bk}
\end{equation}
%%--
where $\beta$ denotes a scaling factor to transform the uplink CSI feedback rate into bandwidth. And the downlink bandwidth can be expressed as
%%--
\begin{equation}
    B_{DL}=B - B_{UL} = B-\beta\sum_{k=1}^{N_s}r_k.
    \label{Eq:Bdl}
\end{equation}
%%--
%%%%%%%%%%%%%%%%%%%%%%%%%%%%%%%%%%%%%%%%%%%%%%%%%%%%%%%%%%%%%%%%%%%
\subsubsection{Carrier Sense Multiple Access}
%%%%%%%%%%%%%%%%%%%%%%%%%%%%%%%%%%%%%%%%%%%%%%%%%%%%%%%%%%%%%%%%%%%
%%%%--------------
Likewise, the BS serves MSs simultaneously in the same frequency band, while each MS feeds back the CSI through Carrier Sense Multiple Access (CSMA), which can listen to channel before transmitting a packet to avoid the avoidable collisions. Sender retransmits after some random time if there is a collision. For efficiency, slotted CSMA is considered:  Time is slotted and a packet can only be transmitted at the beginning of one slot. Sender finds out whether transmission was successful or experienced a collision by listening to the ACK/NACK broadcast from the receiver.

Without loss of generality, we consider the slotted $p$-persistent CSMA in \cite{P-P}, which can be described by the following steps:
%%%%--------
\begin{itemize}
  \item If the channel is idle, transmit with probability $p$, and delay for worst case propagation delay for one packet with probability $1-p$;
  \item If the channel is busy, continue to listen until medium becomes idle, then go to Step $1$;
  \item If transmission is delayed by one time slot, continue with Step $1$.
\end{itemize}
%%%%--------
For slotted $p$-persistent CSMA, the throughput ($S$) is given by \cite{P-P}:
%%%%--------
\begin{align}
S=&\frac{G\sum^{\infty}_{k=0}(p(1-p)^{k}+\alpha[1-(1-p)^{k+1}])\cdot\exp\big(G(1-p)^{k+1}+\alpha G[-(k+1)+\frac{1-(1-p)^{k+2}}{p}]\big)}{(1+\alpha)\cdot\exp(G(1+\alpha))+\alpha \sum_{k=1}^{\infty}\cdot\exp\big(G(1-p)^k+\alpha G[-k+\frac{1-(1-p)^{k+1}}{p}]\big)},
\label{Eq:ppersist}
\end{align}
%%%%--------
%%%%%--------
%\begin{equation}
%S=\frac{G\sum^{\infty}_{k=0}(p(1-p)^{k}+\alpha[1-(1-p)^{k+1}])\cdot\exp\big(G(1-p)^{k+1}+\alpha G[-(k+1)+\frac{1-(1-p)^{k+2}}{p}]\big)}{(1+\alpha)\cdot\exp(G(1+\alpha))+\alpha \sum_{k=1}^{\infty}\cdot\exp\big(G(1-p)^k+\alpha G[-k+\frac{1-(1-p)^{k+1}}{p}]\big)},
%\end{equation}
%%%%%--------
where $\alpha=\frac{\tau}{T}$, $\tau$ is the propagation delay, $T$ is the packet transmission time, and $G$ is the offered load~(overall rate). By CSMA, each user tries to adjust its requested feedback rate $r_k$ over the uplink bandwidth $B_{DL}$. However, if the overall rate is too high, due to the random access nature, the network would be congested. As a result, the accurate rate $z_k$ of user $k$ will reduce a lot. From \cite{data_network}, we have
\begin{eqnarray}\label{Eq:CSMArate}
z_k(r_k, \textbf r_{-k})=\left\{
\begin{array}{ll}
\frac{r_kS}{G}, & \mbox{if } G\leq G_0,\\
0, & \mbox{otherwise,}
\end{array}
\right.
\end{eqnarray}
where the overall rate $G=\sum _{k} r_{k}$, and $G_0$ is the maximum network payload. Similarly, the downlink bandwidth can be calculated as that in (\ref{Eq:Bdl}): $B_{DL}= B-\beta\sum_{k=1}^{N_s}r_k$.
%%%%%%%%%%%%%%%%%%%%%%%%%%%%%%%%%%%%%%%%%%%%%%%%%%%%%%%%%%%%%%%%%%%
\subsection{CSI Feedback Rate Model}
%%%%%%%%%%%%%%%%%%%%%%%%%%%%%%%%%%%%%%%%%%%%%%%%%%%%%%%%%%%%%%%%%%%
In a closed-loop wireless communication system, the MS needs to feed back the quantized CSI back to the BS to perform transmit precoding. For simplicity and without loss of generality, we here use the equivalent \emph{quantized feedback channel} by transforming the real channel matrix in terms of feedback rate and distortion. We consider a limited and lossless feedback channel. Through CSI quantization, the real channel output for the $k$-th MS, denoted by $\textbf{h}_{k}$, can be modeled as \cite{Zhang07}
%%--
\begin{equation}
    \mathbf{h}_{k}=\overline{\mathbf{h}}_{k}+\mathbf{n}_{s},
    \label{Eq:hk}
\end{equation}
%%--
where $\overline{\mathbf{h}}_{k}\in\mathbb{C}^{N_t{\times}1}$ represents the quantized feedback channel output with zero mean and $1-D_k$ variance, $\mathbf{n}_{s}\in\mathbb{C}^{N_t{\times}1}$ is an independent additive noise matrix with each entry corresponding to an i.i.d. Gaussian variable with distribution $\mathcal{CN}(0,D_k)$, and $D_k$ represents the channel quantization distortion constraint. Note that $\overline{\mathbf{h}}_{k}$ and $\mathbf{n}_{s}$ are mutually independent. Due to imperfection in the feedback channel, the quality of the feedback information can be measured by the distortion on the source $\textbf{h}_{k}$ from its representation $\overline{\textbf{h}}_{k}$, which is defined by
%%--
\begin{equation}
    D_k=\|\mathbf{h}_{k}-\overline{\mathbf{h}}_{k}\|^2.
\label{Eq:dis}
\end{equation}
%%--
\begin{lemma}
Given distortion rate $D_k$, the quantized CSI can be modeled as
%%--
\begin{equation}
    \overline{\mathbf{h}}_{k}=\mu\mathbf{h}_{k}+\nu\mathbf{n}_{q},
\label{Eq:qua}
\end{equation}
%%--
where $\mu=1-D_k$, the elements of $\mathbf{n}_{q}\in\mathbb{C}^{N_t{\times}1}$ are i.i.d. Gaussian variables with distribution $\mathcal{CN}(0,1)$, and $\nu=\sqrt{D_k(1-D_k)}$. The detailed derivations of $\mu$ and $\nu$ is given in Appendix~A.
\end{lemma}

Based on the Shannon's rate-distortion theory of continuous-amplitude sources, the rate-distortion function of a zero-mean and unit variance complex Gaussian source is given by~\cite{Proakis1995}
%%--
\begin{equation}
    r_k=\log_2\left(\frac{1}{D_k}\right),
\label{Eq:rat}
\end{equation}
%%--
where $r_k$ represents the feedback rate of user $k$. It can be observed in (\ref{Eq:rat}) that when $D_k=1$, i.e., completely distorted, the feedback rate is equal to zero, while $D_k\rightarrow{0}$, this requires the infinite feedback rate to realize the undistorted CSI.

Substituting (\ref{Eq:rat}) into (\ref{Eq:qua}), the quantized CSI matrix, $\overline{\mathbf{h}}_{k}$, can be expressed as a function of the feedback rate, $r_k$
%%--
\begin{equation}
    \overline{\mathbf{h}}_{k}=(1-2^{-r_k})\mathbf{h}_{k}+\sqrt{2^{-r_k}(1-2^{-r_k})}\mathbf{n}_{q},
\label{Eq:quaH}
\end{equation}
%%--
which clearly connects the feedback rate, $r_k$, with the quantized CSI, $\mathbf{h}$, in order to reveal its impact on system performance. After normalization, (\ref{Eq:quaH}) becomes
%%--
\begin{equation}
    \overline{\mathbf{h}}_{k}=\sqrt{1-2^{-r_k}}\mathbf{h}_{k}+\sqrt{2^{-r_k}}\mathbf{n}_{q},
\label{Eq:Norqua}
\end{equation}
%%--
which can be used to perform precoding at the BS side.

%%%%%%%%%%%%%%%%%%%%%%%%%%%%%%%%%%%%%%%%%%%%%%%%%%%%%%%%%%%%%%%%%%%
\section{Transmit Precoding with Limited CSI Feedback}
%%%%%%%%%%%%%%%%%%%%%%%%%%%%%%%%%%%%%%%%%%%%%%%%%%%%%%%%%%%%%%%%%%%
In this section, we first present transmit precoder implemented at BS, and then discuss a few properties of limited CSI feedback.
%%%%%%%%%%%%%%%%%%%%%%%%%%%%%%%%%%%%%%%%%%%%%%%%%%%%%%%%%%%%%%%%%%%
\subsection{Minimum Mean Square Error Precoder}
%%%%%%%%%%%%%%%%%%%%%%%%%%%%%%%%%%%%%%%%%%%%%%%%%%%%%%%%%%%%%%%%%%%
For simplicity and without loss of generality, we just consider the conventional MMSE based precoder design, and other advanced precoding approaches can be readily applied in this paper. The received signals in (\ref{Eq:Sui}) can be rewritten in a matrix form
%%--
\begin{equation}
    \mathbf{y}=\mathbf{H}\mathbf{W}\mathbf{x}+\mathbf{n},
    \label{Eq:SuiM}
\end{equation}
%%--
where $\mathbf{y}=[y_1,\ldots,y_{N_s}]^T\in\mathbb{C}^{N_s{\times}1}$, $\mathbf{H}=[\mathbf{h}_1,\ldots,\mathbf{h}_{N_s}]^T\in\mathbb{C}^{N_s{\times}N_t}$, $\mathbf{W}=[\mathbf{w}_1,\ldots,\mathbf{w}_{N_t}]\in\mathbb{C}^{N_t{\times}N_t}$, and $\textbf{n}=[n_1,\ldots,n_{N_s}]^T\in\mathbb{C}^{N_s{\times}1}$.

In this section, we demonstrate that the following form for the precoder
%%--
\begin{equation}
    \mathbf{W}=K\overline{\mathbf{H}}^H(\overline{\mathbf{H}}\ \overline{\mathbf{H}}^H+\psi\mathbf{I})^{-1},
    \label{Eq:UiPreGe}
\end{equation}
%%--
with two free scalar parameters $K$ and $\psi$, is a general form for an optimal linear precoder, and $\overline{\mathbf{H}}=[\overline{\mathbf{h}}_1,\ldots,\overline{\mathbf{h}}_{N_s}]^T\in\mathbb{C}^{N_s{\times}N_t}$. By varying the choice of these parameters, optimality can be achieved with respect to a variety of criteria that have been considered in the literature. In general, $K$ is a normalization constant used to comply with the unit transmit power constraint (averaged over data symbols), and it can be expressed as
%%--
\begin{equation}
    K=\parallel\mathbf{T}\parallel^{-1},
    \label{Eq:KPre}
\end{equation}
%%--
where $ \mathbf{T}=\overline{\mathbf{H}}^H(\overline{\mathbf{H}}\ \overline{\mathbf{H}}^H+\psi\mathbf{I})^{-1}$ represents the unnormalized transmit precoder. The other parameter, $\psi$, is typically a regularization parameter. Maximizing the average SINR under flat fading, which is the same for all MSs due to the assumption of symmetry in the distribution of the channel matrix, and taking $N_t$ and $N_s$ (which in this context, is the number of transmit antennas as well as the number of MSs with single antenna) large, the optimal $\psi$ can be expressed as the following general form~\cite{Peel2005}
%%--
\begin{equation}
    \psi \approx {\mathrm{SINR}}^{-1}.
    \label{Eq:psi}
\end{equation}
%%--
Note that although $\psi$ was solved for the equal SINR case in~\cite{Peel2005}, $\psi$ is really a tunable parameter that can be optimized for other criteria. By setting $\psi=0$, (\ref{Eq:UiPreGe}) becomes the simplest and the most common zero-forcing (channel inversion) precoder
%%--
\begin{equation}
    \mathbf{W}=K\overline{\mathbf{H}}^H(\overline{\mathbf{H}}\ \overline{\mathbf{H}}^H)^{-1}.
    \label{Eq:UiPre}
\end{equation}
%%--

Recalling (\ref{Eq:Sui}), the signal to noise plus interference ratio~(SINR) of the $k$-th MS can be written as
%%--
\begin{equation}
    \gamma_k
    =\frac{\mid\mathbf{h}_{k}^T\mathbf{w}_k\mid^2}{\mid\mathbf{h}_{k}^T\sum_{i=1,i{\neq}k}^{N_s}\mathbf{w}_i\mid^2+N_0}.
    \label{Eq:SNRk}
\end{equation}
%%--
Its corresponding throughput can be expressed as
%%--
\begin{equation}
    C_k(\gamma_k)
    =B_{DL}\log_2(1+\gamma_k).
    \label{Eq:ratek}
\end{equation}
%%--
%%%%%%%%%%%%%%%%%%%%%%%%%%%%%%%%%%%%%%%%%%%%%%%%%%%%%%%%%%%%%%%%%%%
\subsection{Analysis of the CSI feedback impact}
%%%%%%%%%%%%%%%%%%%%%%%%%%%%%%%%%%%%%%%%%%%%%%%%%%%%%%%%%%%%%%%%%%%
Since the precoder $\mathbf{W}$ is designed by the feedback CSI, $\overline{\mathbf{H}}$, optimal performance using the MMSE rule can be achieved when $r_k\rightarrow{+\infty}$, for $\forall{i}\in [1,\ldots,N_s]$. We in this subsection present a few properties of (\ref{Eq:SNRk}) based on the CSI feedback:
\begin{enumerate}
  \item $\gamma_k$ is continuous and monotonic increasing in $r_k$.
  \begin{proof}
  It is obvious in (\ref{Eq:SNRk}) that $\gamma_k$ is continuous in $r_k$. In addition, from (\ref{Eq:quaH}), we can see that the increase of $r_k$, i.e. the decrease of $D_k$, improves the amount of the feedback CSI, $\overline{\mathbf{h}}_{k}$. This enhances the accuracy of the constructed precoder, $\mathbf{w}_k$, and thus, increases the value of $\mathrm{SINR}_k$.
  \end{proof}
  \item $\underset{r_k\rightarrow{+\infty}}\lim\gamma_k=\varepsilon_k$, where $\varepsilon_k$ is a constant.
  \begin{proof}
  Given $r_i$ (for $\forall{i}\in [1,\ldots,N_s]$, but $i\neq{k}$), we have $\underset{r_k\rightarrow{+\infty}}\lim\overline{\mathbf{h}}_{k}={\mathbf{h}}_{k}$. Hence, $\gamma_k$ converges to a constant when the feedback rate is large enough.
  \end{proof}
\end{enumerate}

In 1), it implies that the feedback rate ($r_k$) of every MS should be as large as possible. However, if using orthogonal channels in Subsection-II-C-1), this will reduce the downlink bandwidth and then decrease the system throughput. By adopting the CSMA protocol in Subsection-II-C-2), this will results in more collisions in the uplink, and degrade the effective feedback rate. Obviously, there exists a trade-off between feedback rate ($r_k$) and system throughput ($C_k(\gamma_k)$) in both uplink multiple access protocols. Notice that these properties will be reused to prove the existence of the NFCP equilibrium in~Subsection-IV-B.

%%%%%%%%%%%%%%%%%%%%%%%%%%%%%%%%%%%%%%%%%%%%%%%%%%%%%%%%%%%%%%%%%%%
\section{Noncooperative Feedback Control Game for Channel State Information}
%%%%%%%%%%%%%%%%%%%%%%%%%%%%%%%%%%%%%%%%%%%%%%%%%%%%%%%%%%%%%%%%%%%
In this section, we first define the utility function. Then, we describe the noncooperative CSI feedback control game. Next, we use the pricing method to improve the performance of the proposed game. the convergence to the Nash equilibrium is also proved. Finally, we construct a centralized solution for performance comparison.

%%%%%%%%%%%%%%%%%%%%%%%%%%%%%%%%%%%%%%%%%%%%%%%%%%%%%%%%%%%%%%%%%%%
\subsection{Utility Function}
%%%%%%%%%%%%%%%%%%%%%%%%%%%%%%%%%%%%%%%%%%%%%%%%%%%%%%%%%%%%%%%%%%%
The concept of utility is commonly used in microeconomics and refers to the level of satisfaction the decision-taker receives as a result of its actions. MSs access a wireless system through the air interface that is a common resource and they transmit information expending bandwidth resources. Since the air interface is a shared medium, each MS' transmission is a source of competition for others. SINR is an effective measure of the quality of signal reception for the wireless user~\cite{Proakis1995}. An MS tries to achieve a high quality of reception (throughput in this paper) while at the same time expending a certain amount of system bandwidth to feedback CSI. Obviously, the CSI feedback rate determines the accuracy of the precoder, and thus, affect the system throughput. Therefore, it is possible to view both throughput and the CSI feedback rate as commodities that a wireless user desires. The utility function of the $k$-th MS can be expressed as
\begin{equation}
    u_k = C_k(\gamma_k)
        = B_{DL}\log_2(1+\gamma_k).
    \label{Eq:Uk}
\end{equation}
%%--
Utility as defined above is the throughput conditioned on the feedback bit. Note that if $r_k=0$, omnidirectional transmission instead of precoding should be used which results in minimum value of $C_k$. This suggests that, in order to maximize utility, all users in the system should feed back a certain amount of CSI. For orthogonal feedback channels in (\ref{Eq:Bdl}), when the amount of feedback increases, downlink bandwidth, $B_{DL}$, will decrease, and thus reduce the system throughput. With regard to CSMA, when the network payload increases, more collisions happen and consequently the average delay for each packet increases. Any network payload larger than $G_0$ in (\ref{Eq:CSMArate}) will cause an unacceptable average delay. As a result, the utility becomes zero. All these facts indicate that the feedback rate should not be either too small or too large for better utility. In other word, there is an optimal point on how much to feedback from each MS point of view.

\begin{figure}[]
\centering
\includegraphics[height=3.6in,width=4.5in]{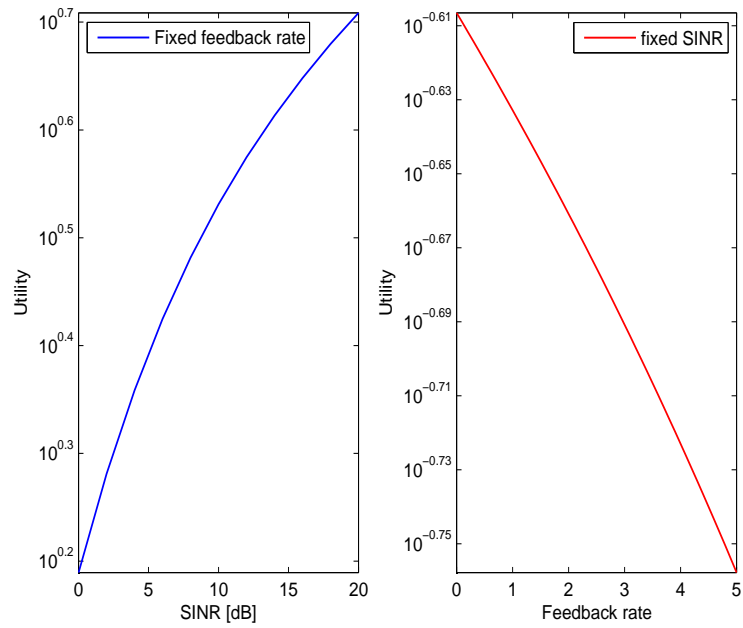}
\caption{The behavior of the MS' utility as a function of SINR for fixed feedback rate, and as a function of feedback rate for fixed SINR.}
\label{Fig:Utility_tradeoff}
\end{figure}

Intuitively, there exists a tradeoff relationship between obtaining high throughput and requiring small amount of CSI feedback on the condition of a total system bandwidth constraint. Finding a good balance between the two conflicting objectives is the primary focus of the CSI feedback rate control component of radio resource management. This tradeoff is illustrated through the conceptual plot in Fig.~\ref{Fig:Utility_tradeoff}, where orthogonal feedback channel is assumed. If the feedback rate were fixed, the terminal would experience higher throughput as the SINR increases which leads to increased satisfaction of the use of the system resources. If the SINR were to be fixed (fixed throughput), increasing the feedback rate expedites uplink bandwidth, which effectively reduces the satisfaction of the mobile terminal. For sufficiently large SINR values, the throughput approaches zero, which results in an asymptotic decrease in utility in the high SINR region.

%%%%--------------
\subsection{Game Formulation}
%%%%--------------
In the sequel, we first consider the noncooperative feedback rate control game (NFC) where each MS tries to maximize its individual utility. Let $\mathcal{G}=[N,\{\mathcal{R}_k\},\{u_k(\cdot)\}]$ denote the NFC where $\mathcal{N}=\{1,\ldots,N_s\}$ is the index set for the mobile users currently in the cell, $\mathcal{R}_k$ is the strategy set, and $u_k(\cdot)$ is the payoff function of user $k$. Each user selects a feedback rate level $r_k$ such that $r_k\in\mathcal{R}_k$. Let the feedback rate vector $\mathbf{r}=(r_1,\ldots,r_{N_s})\in\mathcal{R}$ denote the user's strategies in terms of the selected rate levels of all the users, where $\mathcal{R}$ is the set of all rate vectors. The resulting utility level for the $k$-th user is $u_k(r_k,\mathbf{r}_{-k})$, where $\mathbf{r}_{-k}$ denotes the vector consisting of the other user's strategies other than the $k$-th user. This notation emphasizes that the $k$-th user has control over its own rate, $r_k$ only. The utility of the $k$-th MS with feedback rate $r_k$ can be expressed more rigorously as
%%--
\begin{equation}
    u_k(r_k,\mathbf{r}_{-k}) = B_{DL}\log_2(1+\gamma_k(r_k,\mathbf{r}_{-k})).
    \label{Eq:Ukr}
\end{equation}
%%--
Note that (\ref{Eq:Ukr}) demonstrates the strategic interdependence between MSs. The level of utility each MS gets depends on its own feedback rate and also on the choice of other players' strategies, through the SINR $\gamma_k$ of that user. The efficiency function can be chosen to represent any precoding scheme described in Section III. In this paper, we assume that the strategy space, $\mathcal{R}_k$, of each user is a compact, convex set with minimum and maximum rate constraints denoted by $r_k^{\min}$ and $r_k^{\max}$, respectively. For simplicity, we let $r_k^{\min}=0$ for all $k$, which results in the strategy space $\mathcal{R}_k=[0,r_k^{\max}]$. The utility function takes the generic form given in Fig. 3 for fixed interference plus noise.

The NFC game can be expressed as
%%--
\begin{align}
    (\mathrm{NFC})\ {\max_{r_k\in{R_k}}}\ u_k(r_k,\mathbf{r}_{-k}), \ \forall k\in\mathcal{N}.
    \label{Eq:NFC}
\end{align}
%%--
From (\ref{Eq:NFC}), it clearly indicates that the feedback rate that optimizes individual utility depends on rates of all the other MSs in the network. It is necessary to characterize a set of rates where the users are satisfied with the utility they receive given the rate selections of other users.

NFC offers a solution to the rate control problem where no MS can increase its utility any further through individual effort. Thus, it is an outcome obtained as a result of distributed decision taking, which could be expected to be less efficient than a possible rate selection obtained through cooperation between terminals and/or as a result of centralized optimization. Specifically, in the NFC, each MS aims to maximize its own utility by adjusting its own feedback rate, but it ignores the cost (or harm) it imposes on the other terminals. For example, by orthogonal feedback channels, if one MS increases the usage of bandwidth in order to send more CSI, it will decrease the available bandwidth for other MSs. While for CSMA, the higher feedback rate will cause the heavier collision possibility.

\subsection{Pricing Mechanism}

To overcome this problem, we resort to a usage-based pricing schemes. By introducing a pricing factor for the feedback CSI, we can increase system performance by implicitly inducing cooperation, and yet we maintain the noncooperative nature of the resulting feedback rate of the CSI control solution. Within the context of a resource allocation problem for a closed-loop wireless system, the resource being shared is the radio environment, and the resource usage is determined by MS's feedback rate. Hence, efficiency in feedback rate control can be promoted by the proposed usage-based pricing strategy where each user pays a penalty proportional to its usage amount, i.e. rate, of feedback CSI.

The NFCP can be expressed as the following optimization problem
%%--
\begin{align}
    (\mathrm{NFCP})\ {\max_{r_k\in{R_k}}}\ u_k^c(r_k,\mathbf{r}_{-k})
    =u_k(\mathbf{r}_k)-c_k(r_k), \ \forall k\in\mathcal{N}
    \label{Eq:mo}
\end{align}
%%--
where $\mathcal{G}_c=[N,\{\mathcal{R}_k\},\{u_k^c(\cdot)\}]$ represents a $N_s$ player noncooperative feedback rate control game with pricing~(NFCP),  $u_k^c(r_k,\mathbf{r}_{-k})$ is the utility for NFCP, and $c_k(r_k)$ denotes the pricing function for the $k$-th MS, which in this paper is restricted to linear schemes of the form
%%--
\begin{equation}
    c_k(r_k)=\alpha r_k.
    \label{Eq:pricef}
\end{equation}
%%--
Here $\alpha$ is  announced by the BS and is a constant referred to as the price factor per feedback bandwidth.

Note that (\ref{Eq:mo}) also demonstrates the strategic interdependence between users, and the pricing factor needs to be tuned such that user self-interest leads to the best possible improvement in overall network performance. Combining (\ref{Eq:Uk}) and (\ref{Eq:pricef}), the NFCP with linear price in (\ref{Eq:mo}) is as follows
%%--
\begin{align}
    (\mathrm{NFCP})\ {\max_{r_k\in{R_k}}}\ u_k^c(r_k)
    =B_{DL}\log_2(1+\gamma_k)-\alpha r_k, \ \forall k\in\mathcal{N}.
    \label{Eq:mof}
\end{align}
%%--

\begin{table}[]
\caption{NFCP algorithm for each ms} \label{Tab: NFCP_terminal} \centering
\begin{tabular}{p{100mm}}
\hline \textbf{Algorithm 1: Non-cooperative CSI feedback rate control game with a given price $\alpha$ at the MS side} \\
\hline \quad 1. Set initial CSI vector at time $t=0$: $\textbf{r}(0)=\textbf{r}_0$. Also, let $k=1$;

\quad 2. For all $j$, such that $\tau_j\in T$:

\quad \quad $\ast$  Given $\textbf{r}_{-k}(\tau_{j-1})$, compute: \\
\quad \quad\quad \quad \quad $r_k(\tau_j)=\texttt{arg}\max_{r_k \in \textbf{r}_k } u_k^c(r_k,\mathbf{r}_{-k}(\tau_{j-1})$.
%\quad \quad $\ast$  Assign the CSI rate as $r_k(\tau_j) = \min (r_k(\tau_j)) $.
\\

\hline
\end{tabular}
\end{table}

By considering the NFCP algorithm in (\ref{Eq:mof}), given $\alpha$, a sequence of rates can be generated in Table~\ref{Tab: NFCP_terminal}. We refer to $r_k(\tau_j) $ as the \emph{set} of best feedback rates for the $k$-th MS at time $j$ instance in response to the interference vector $\textbf{r}_{-k}(\tau_{j-1})$. %Note that in the game with price, more than one feedback rate might constitute a best response to a given interference vector. In this case, the algorithm determines the CSI rate of an MS by selecting the smallest rate among all possibilities as dictated by the algorithm.
For the network level algorithm for each value of $\alpha$, we may first run the NFCP when $\alpha=0$, which is equivalent to the NFC described in (\ref{Eq:NFC}). Once the equilibrium with no price is obtained, the NFCP is played again after incrementing the price factor, $\alpha$, by a positive value, $\triangle\alpha$. Algorithm \ref{Tab: NFCP_terminal} returns a set of CSI rates at equilibrium with this value of the price factor. If the utilities at this new equilibrium with some positive price improve with respect to the previous instance, the price factor is incremented and the procedure is repeated. We continue until an increase in results in utility levels worse than the previous equilibrium values for at least one user. We declare the last value to be the best price factor, $\alpha_{\texttt{BEST}}$. The way that $\alpha_{\texttt{BEST}}$ is determined by the network is summarized in algorithmic format in Table~\ref{Tab: NFCP_Network}.

%%%%--------------
\begin{table}[ht]
\renewcommand{\arraystretch}{1.3}
\caption{NFCP algorithm for the network}\label{Tab: NFCP_Network} \centering
\begin{tabular}{p{100mm}}
\hline \textbf{Algorithm 2: Non-cooperative CSI feedback rate control game with price algorithm for the whole network} \\
\hline \quad 1. Set $\alpha=0$ and announce $\alpha=0$ to all MSs;

\quad 2. Get $u_k$ for all at equilibrium using $\mathbf{Algorithm \ 1}$, increase $\alpha: = \alpha + \triangle\alpha$, and then announce to all MSs;

\quad 2. If $u_k^{\alpha} < u_k^{\alpha + \delta \alpha}$ for all then go to step $2$, else stop and declare $\alpha_{\texttt{BEST}} = \alpha$.
\\
\hline
\end{tabular}
\end{table}

%%%%--------------
\subsection{Nash Equilibrium}
%%%%--------------
In this subsection, we investigate the equilibrium of the proposed games, at which no player can improve its utility by changing its own strategy only. Note that since NFC can be treated as a special case of NFCP when $\alpha=0$, it would be sufficient to declare the equilibrium existence of the NFC if a Nash equilibrium exists in the NFCP.

\begin{definition}

A rate vector $r=(r_1,\ldots,r_{N_s})$ is a Nash equilibrium of the NPCP $\mathcal{G}=[N,\{\mathcal{R}_k\},\{u_k^c(\cdot)\}]$ if, for every $k\in\mathcal{N}$, $u_k^c(r_k,\mathbf{r}_{-k})\geq u_k^c(r'_k,\mathbf{r}_{-k})$, $r'_k\in\mathcal{R}_k$.
\end{definition}

There are some existing theorems to show the existence. We only need to prove that the proposed game satisfies the requirements of the theorems. It has been shown a Nash equilibrium exists, for $\forall \ k$:

\begin{theorem}
A Nash equilibrium exists in the NFCP, $G=[N, \{r_k\}, u_k^c(\cdot)]$ if $\forall k \in N$:
\end{theorem}

\begin{enumerate}
  \item $\mathbf{r}_k$, the support domain of $u_k(\mathbf{r}_k)$, is a nonempty, convex, and compact
subset of a certain Euclidean space $\mathbf{R}$.
  \item $u_k(\mathbf{r}_k)$ is continuous in $\mathbf{r}$ and quasi-concave in $\mathbf{r}_k$.
\end{enumerate}

\begin{proof}
Obviously, the support domain $\mathbf{r}_k$, which is a vector, satisfies the first condition.

To prove that $u_k^c(\mathbf{r}_k)$ is quasi-concave, it is equivalent to prove that the first-order derivative of $u_k^c(\mathbf{r}_k)$ is a monotonic decreasing function whose value varies from positive to negative in term of $\mathbf{r}_k$~\cite{Gradshteyn94}. For convenience, let $b_k(\mathbf{r}_k)\triangleq{B-\beta\sum_{k=1}^{N_s}r_k}$ and $w_k(\mathbf{r}_k)\triangleq\log_2(1+\gamma_k)$, and thus $u^c_k(\mathbf{r}_k)=b_k(\mathbf{r}_k)w_k(\mathbf{r}_k)-\alpha r_k$. The first-order derivative of $u_k^c(\mathbf{r}_k)$ with respect to $r_k$ can be calculated as

%%--
\begin{align}
    u^c_k{'}(\mathbf{r}_k)=b_k'(\mathbf{r}_k)w_k(\mathbf{r}_k)+b_k(\mathbf{r}_k)w_k'(\mathbf{r}_k)-\alpha,
    \label{Eq:fird}
\end{align}
%%--
where $b_k'(\mathbf{r}_k)=-\beta$.

As $\gamma_k$ is continuous and monotonic increasing in $r_k$, proved in Subsection-III-C-1, it is obvious that $w_k(\mathbf{r}_k)$ is also continuous and monotonic increasing in $r_k$, i.e. $w_k'(\mathbf{r}_k)>0$. In addition, since $w_k(\mathbf{r}_k)$ is concave in term of $r_k$, $w_k{''}(\mathbf{r}_k)\leq 0$, which indicates that $w_k'(\mathbf{r}_k)$ is a decreasing function. Because $b_k(\mathbf{r}_k)$ is monotonic decreasing in $r_k$, we may conclude that both $b_k'(\mathbf{r}_k)w_k(\mathbf{r}_k)$ and $b_k(\mathbf{r}_k)w_k'(\mathbf{r}_k)$ are decreasing functions. Hence, $u^c_k{'}(\mathbf{r}_k)$ is monotonic decreasing in $r_k$.

As $\underset{r_k\rightarrow{+\infty}}\lim\gamma_k=\varepsilon_k$, we may easily obtain $\underset{r_k\rightarrow{+\infty}}\lim{w_k'(\mathbf{r}_k)=0}$. From ($\ref{Eq:fird}$), given the pricing factor $\alpha$, we can then have:
\begin{itemize}
  \item ${\underset{r_k\rightarrow{0}}\lim}u_k^c{'}(\mathbf{r}_k)=Bw_k'(\mathbf{r}_k)-\alpha>0$;
  \item ${\underset{r_k\rightarrow{+\infty}}\lim}u_k^c{'}(\mathbf{r}_k)={-\beta}w_k(\mathbf{r}_k)-\alpha<0$.
\end{itemize}

Hence, $u_k^c(\mathbf{r}_k)$ is a concave function, and as every concave function is quasiconcave, \emph{Theorem 1} is proved. Finally, we prove the existence of the equilibrium of the game, and can conclude that game $\mathcal{G}$ of (\ref{Eq:mof}) always admits at least one Nash equilibrium.
\end{proof}
\subsection{Centralized Scheme}
%%%%--------------
To compare the performance, a centralized scheme is constructed assuming all CSI is known. The objective is to optimize the sum rate capacity defined in (\ref{Eq:Uk}) subject to the constraint of feedback rates:
\begin{align}
    &\max_{r_i} \sum_{i=1}^{N_s}\ u_i(\mathbf{r}_i)
    %= \max_{r_i} \sum_{i=1}^{N_s} \left(B-\beta\sum_{i=1}^{N_s}r_i\right)\log_2(1+\mathrm{SNR}_i),
    \nonumber \\
    \mathrm{s.t.} \ &B-\beta\sum_{i=1}^{N_s}r_i>0,
    \nonumber \\
    &0\leq{r_i}\leq{r_{\max}}.
    \label{Eq:Centralized}
\end{align}
%%--
Notice that our proposed noncooperative game theoretic based algorithm is distributive, in the sense that only the price information needs to be exchanged, while the centralized scheme needs to gather all the information, which will cause significant signalling.

%%%%%%%%%%%%%%%%%%%%%%%%%%%%%%%%%%%%%%%%%%%%%%%%%%%%%%%%%%%%%%%%%%%
%%%%--------------
\subsection{Implementation Discussion}
%%%%--------------
%%%%%%%%%%%%%%%%%%%%%%%%%%%%%%%%%%%%%%%%%%%%%%%%%%%%%%%%%%%%%%%%%%%
There are several implementation issues for the proposed scheme. Firstly, the channel estimation for the downlink channel might not be accurate due to both fast fading and noise effects. Under this condition, the transmit precoder formula should be rewritten considering the estimation inaccuracy. Secondly, the proposed scheme needs iteratively update the price and rate information. A natural question arises if the distributed scheme has less signalling than the centralized scheme. The comparison is similar to distributed and centralized power control in the literature~\cite{Han2008,Han2010}. Since the channel condition is continuously changing, the distributed solution only needs to update the difference of the parameters such as rate and pricing factor, while the centralized scheme requires all channel information in each time period. As a result, the distributed solution has a clear advantage and dominates the current and future wireless network design. For example, the power control for cellular networks, the open loop power control is done only once during the link initialization, while the close-loop power control (distributed power allocation such as~\cite{Han2008}) is performed $1500$ times for UMTS and $800$ times for CDMA2000.
Finally, for the multi-BS multi-MS case, we can use clustering method to divide the network into sub-networks, and then employ the single BS-MSs solution proposed in this paper.

%%%%%%%%%%%%%%%%%%%%%%%%%%%%%%%%%%%%%%%%%%%%%%%%%%%%%%%%%%%%%%%%%%%
\section{Simulation Results}
%%%%%%%%%%%%%%%%%%%%%%%%%%%%%%%%%%%%%%%%%%%%%%%%%%%%%%%%%%%%%%%%%%%
%%%%%%%%%%%%%%%%%%%%%%%%%%%%%%%%%%%%%%%%%%%%%%%%%%%%%%%%%%%%%%%
In this section, we provide simulation results for the proposed distributed games. All simulations are performed for a BPSK modulation over the Rayleigh fading channels with the MMSE precoder in (\ref{Eq:UiPreGe}). For simplicity, we assume that both the transmit power and the noise variance are normalized to unit. The specific parameters are given below each figure.

%%%%%%%%%%%%%%%%%%%%%%%%%%%%%%%%%%%%%%%%%%%%%%%%%%%%%%%%%%%%%%%%%%%
\subsection{Results through Orthogonal Feedback Channels}
\begin{figure}[]
\centering
\includegraphics[height=3.6in,width=4.5in]{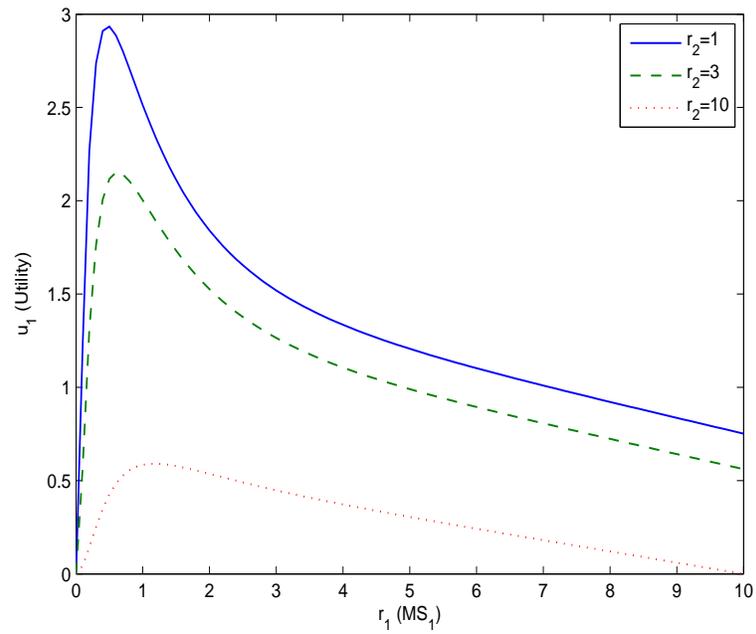}
\caption{The utility of $\mathrm{MS}_1$ in term of $r_1$, where the number of $\mathrm{MSs}$ is $2$ and $r_2 = 1, 3, 10$ over orthogonal feedback channels.}
\label{Fig:NFCP_UE}
\end{figure}

%%%%%%%%%%%%%%%%%%%%%%%%%%%%%%%%%%%%%%%%%%%%%%%%%%%%%%%%%%%%%%%%%%%
In this subsection, we first provide simulations to evaluate the impact of CSI feedback on each individual. Here we plot the utility of MS$_1$ in term of its CSI feedback rate by fixing the feedback rates of the rest MSs. For simplicity, we assume $2$ MSs and the CSI feedback rate of MS$_2$ is $r_2 = 1, 3$, and $10$, respectively. From Fig. \ref{Fig:NFCP_UE}, we can see that when the feedback rates of other users are fixed (fixed uplink bandwidth), the target MS will first experience high throughput as its CSI feedback rate increases, which then leads to increased satisfaction of the use of the system resources. For sufficiently large $r_1$ values, the utility of MS$_1$ begins to decrease. It is obvious that the utility function of each MS is a concave function in terms of the feedback rate, which again partially proves through simulations the existence of equilibrium of the proposed NFCP game.

\begin{figure}[]
\centering
\includegraphics[height=3.6in,width=4.5in]{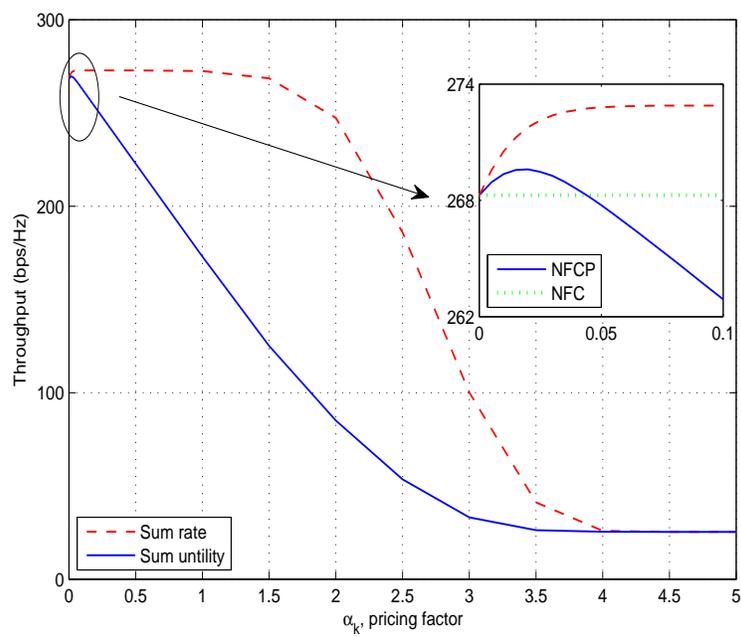}
\caption{Performance of NFCP over orthogonal feedback channels, where the number of MSs is 10, $B=20$, and $\beta=0.01$.}
\label{Fig:NFCP_MS}
\end{figure}

Fig.~\ref{Fig:NFCP_MS} is constructed by letting the algorithm in Table~\ref{Tab: NFCP_terminal} reach the Nash equilibrium at each value of $\alpha$. The best price factor can be found if all mobile users receive worse overall payoff than the previous equilibrium utility according to algorithm~\ref{Tab: NFCP_Network}. It can be also observed from the figure when the pricing factor increases, the total utility and the sum rates first increase, as shown in the small window, and then begin to decrease. It indicates that solution by NFCP with $\alpha = \alpha_{\texttt{BEST}}=0.025$ offers a significant improvement in total utilities with respect to the NFC when $\alpha = 0$, where pricing factor is not involved. At high pricing factor, we can see both sum utility and rate converge to a constant value. This is because the system stops requiring users feedback CSI as it costs too much.

\begin{figure}[]
\centering
\includegraphics[height=3.6in,width=4.5in]{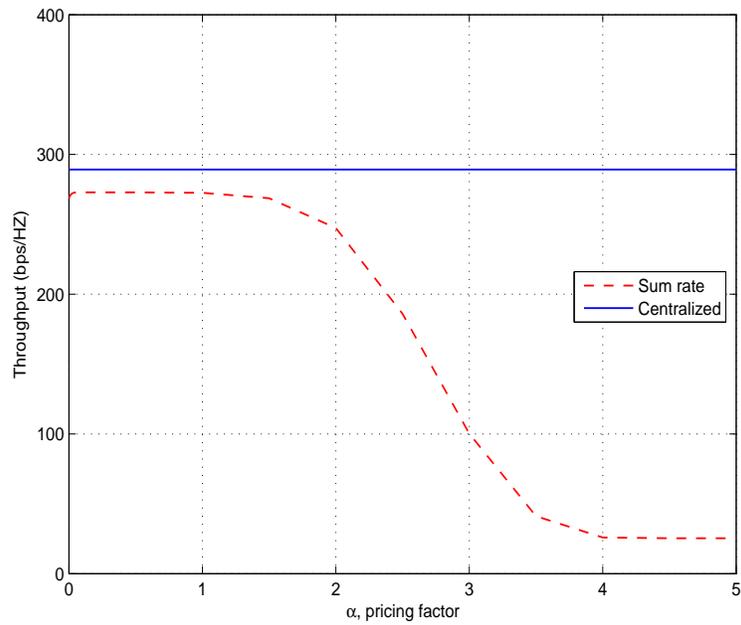}
\caption{Performance comparisons of NFCP and the centralized scheme over orthogonal feedback channels, where the number of MSs is 10, $B=20$, and $\beta=0.01$.}
\label{Fig:NFCP_2MS}
\end{figure}

\begin{figure}[]
\centering
\includegraphics[height=3.6in,width=4.5in]{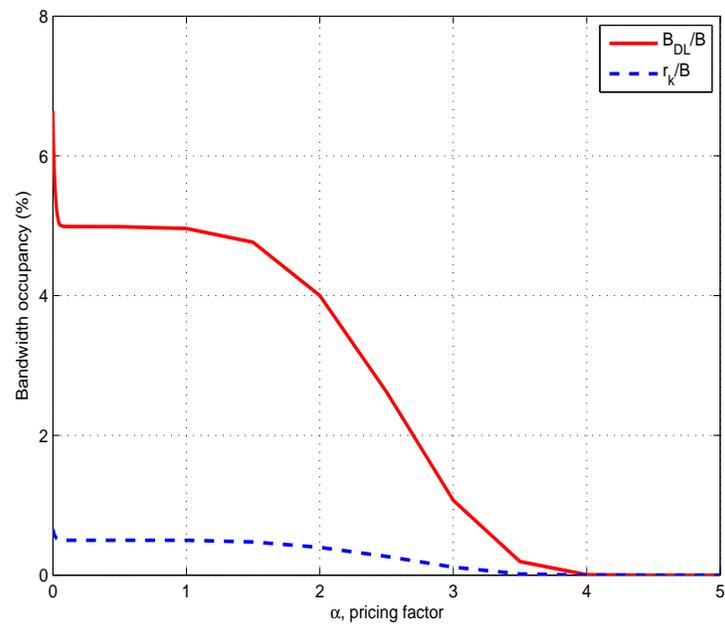}
\caption{Uplink bandwidth occupancy in terms of the pricing factor over orthogonal feedback channels, where the number of MSs is 10, $B=20$, and $\beta=0.01$.}
\label{Fig:NFCP_DL_2MS}
\end{figure}

In Fig.~\ref{Fig:NFCP_2MS}, we compare the proposed NFCP game theoretical approach with the centralized scheme. From the simulation results, we can see that the distributed solution and the centralized solution are asymptotically the same if $\alpha$ is in the right region. When $\alpha$ is too large, the MSs will be reluctant to feedback. When $\alpha$ is too small, the MSs will feed back in a non-cooperative manner. In Fig.~\ref{Fig:NFCP_DL_2MS}, it shows that the variations of the sum feedback rate as well as the individual feedback rate in term of the pricing factor, where $B_{UL} = \beta\sum_{k=1}^{N_s}r_k$. From the figure we can see that these results match very well the sum rate result in Fig.~\ref{Fig:NFCP_MS}. When $\alpha=0$, it requires the maximum amount of feedback. But with the price increase, the feedback rate starts to decrease until zero, which make the throughput dropped to minimum.

%%%%%%%%%%%%%%%%%%%%%%%%%%%%%%%%%%%%%%%%%%%%%%%%%%%%%%%%%%%%%%%%%%%
\subsection{Results through CSMA}
\begin{figure}[]
\centering
\includegraphics[height=3.6in,width=4.5in]{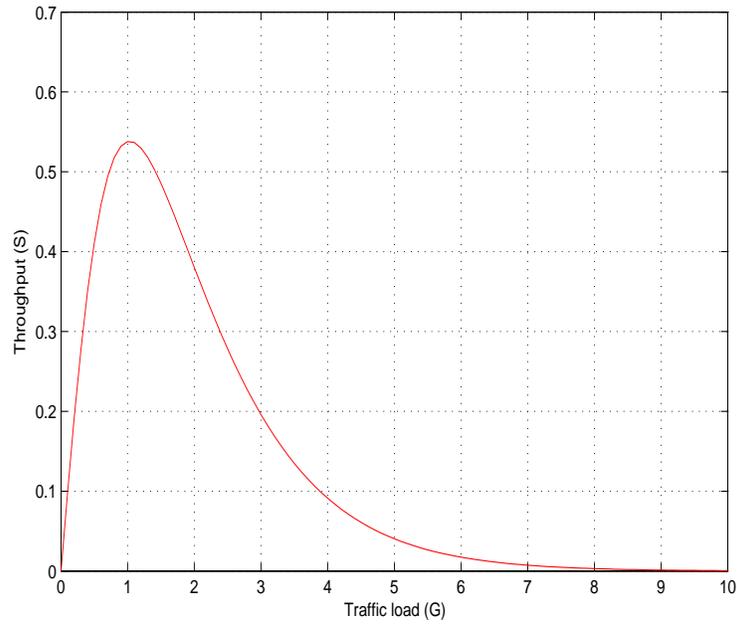}
\caption{Throughput of slotted $1$-persistent CSMA.}
\label{Fig:CSMA}
\end{figure}

\begin{figure}[]
\centering
\includegraphics[height=3.6in,width=4.5in]{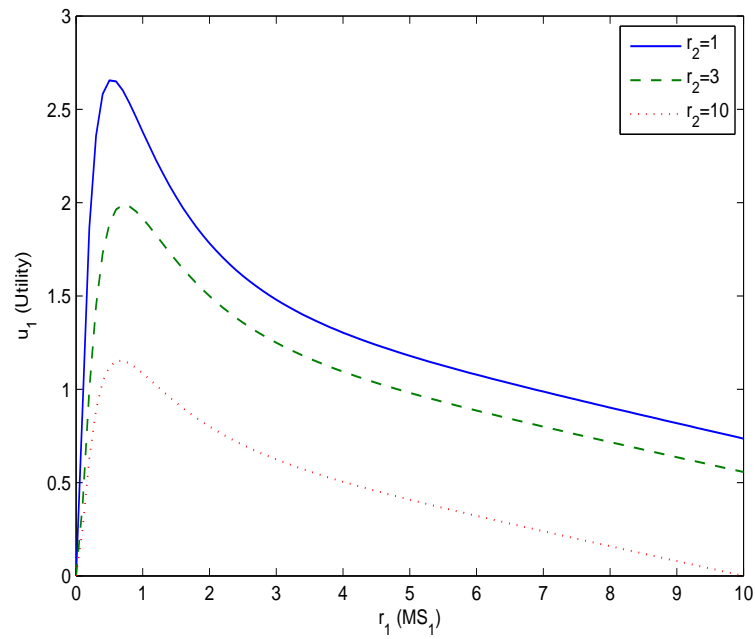}
\caption{The utility of $\mathrm{MS}_1$ in term of $r_1$ over CSMA feedback channels, where the number of $\mathrm{MSs}$ is $2$ and $r_2 = 1, 3, 10$.}
\label{Fig:NFCP_CSMA_UE}
\end{figure}

\begin{figure}[]
\centering
\includegraphics[height=3.6in,width=4.5in]{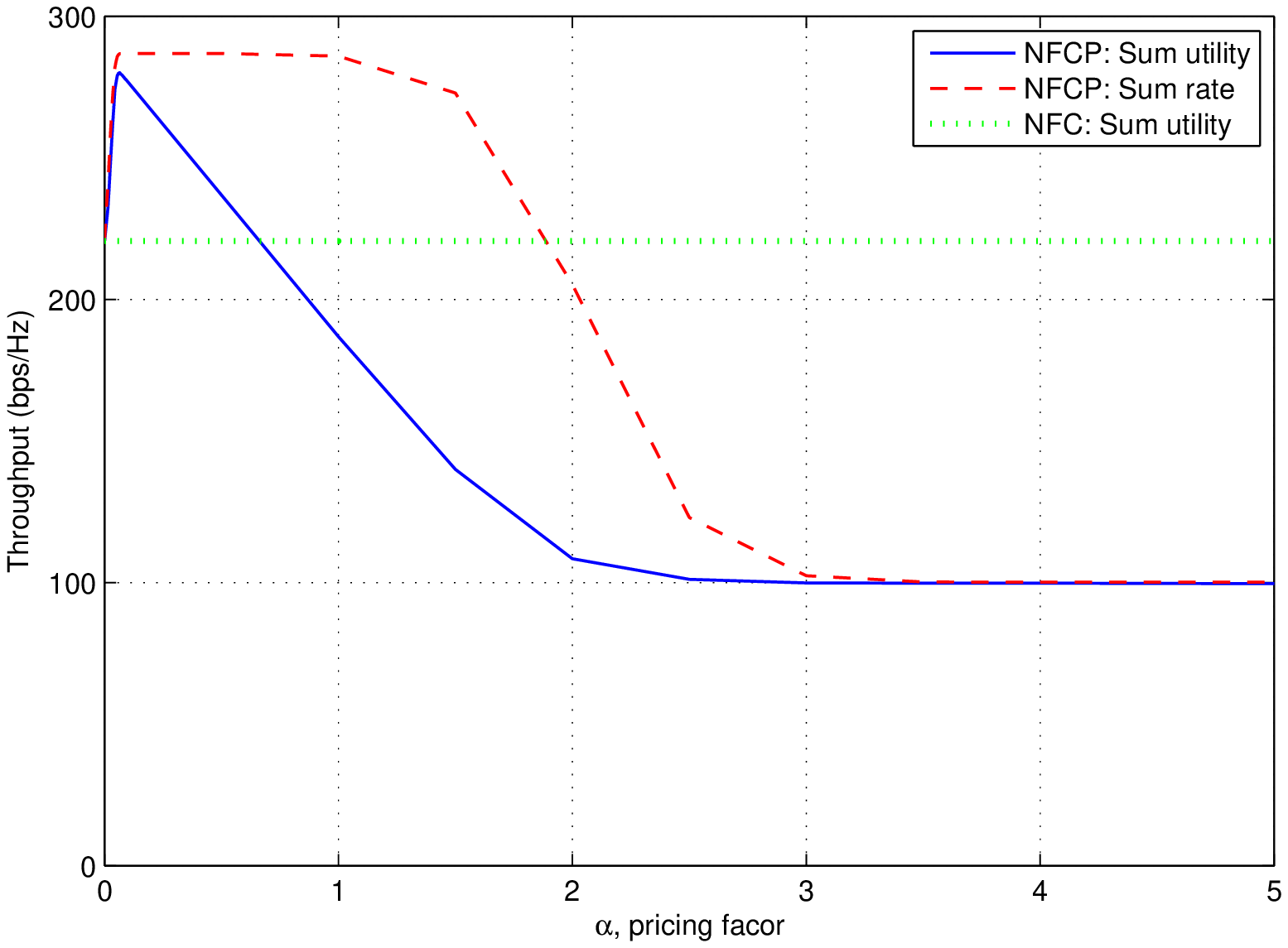}
\caption{Performance of NFCP over CSMA feedback channels, where the number of MSs is 10, and $B=20$.}
\label{Fig:NFCC_CSMA_2MS}
\end{figure}

\begin{figure}[]
\centering
\includegraphics[height=3.6in,width=4.5in]{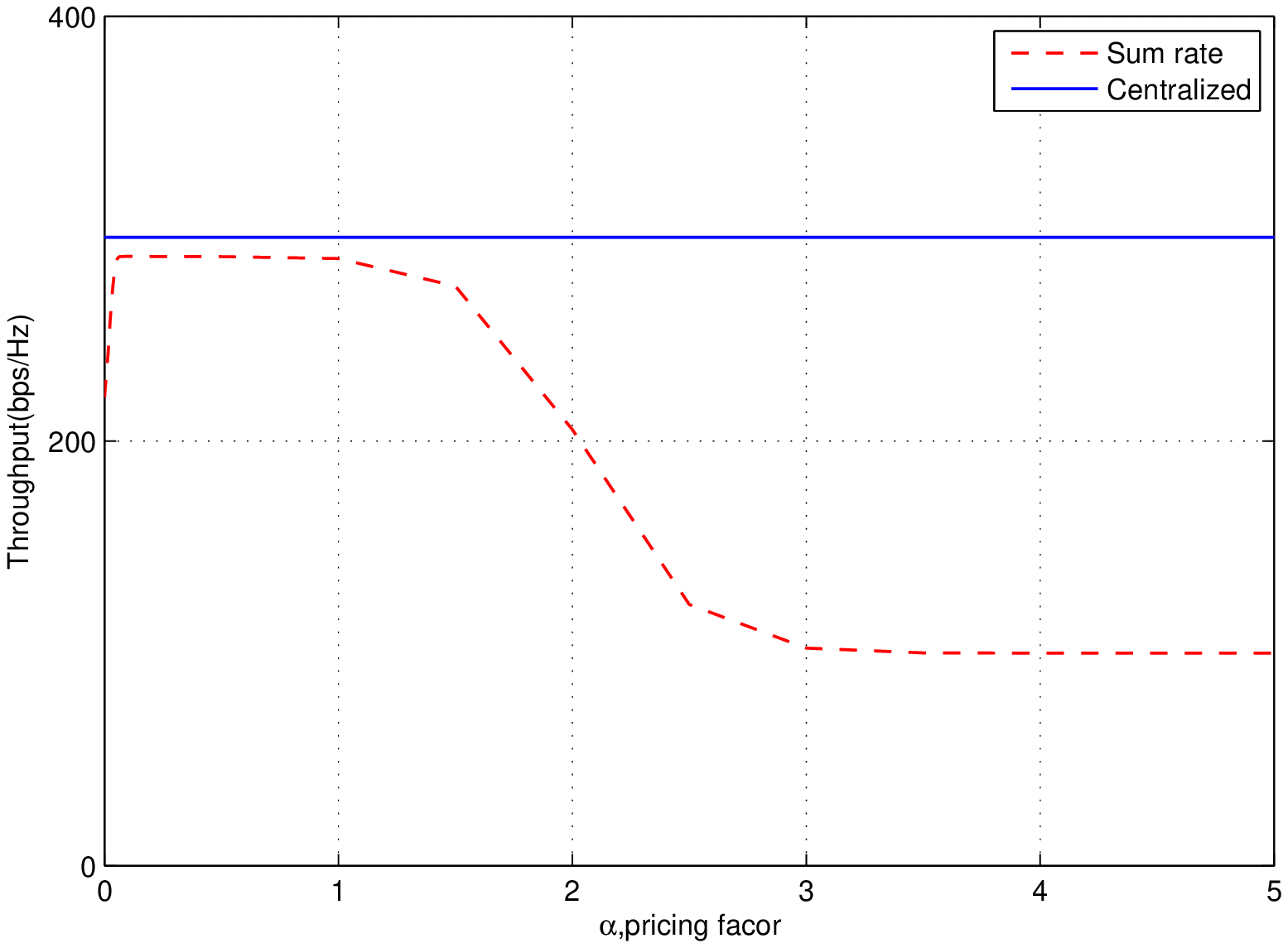}
\caption{Performance of NFCP and the centralized scheme over CSMA feedback channels, where the number of MSs is 10, and $B=20$.}
\label{Fig:NFCC_CSMA_C2MS}
\end{figure}

%%%%%%%%%%%%%%%%%%%%%%%%%%%%%%%%%%%%%%%%%%%%%%%%%%%%%%%%%%%%%%%%%%%
For simplicity, we here consider a special case of slotted $p$-persistent CSMA by setting $p=1$, i.e., $1$-persistent CSMA.
In Fig.~\ref{Fig:CSMA}, it shows the total throughput again the traffic load. It indicates that there exists an optimal transmission rate corresponding to the maximum throughput. In Fig.~\ref{Fig:NFCP_CSMA_UE}, we examine the impact of CSI feedback on each individual using CSMA. From Fig.~\ref{Fig:NFCP_CSMA_UE}, we can see that the utility function of each MS is a concave function in terms of the feedback rate, which again proves the effectiveness of the proposed NFCP game. In Fig.~\ref{Fig:NFCC_CSMA_2MS}, we evaluate the throughput performance in term of the pricing factor. It shows in Fig.~\ref{Fig:NFCC_CSMA_2MS} that the proposed NFCP provides much better results than the NFC game. Fig.~\ref{Fig:NFCC_CSMA_C2MS} compares the proposed NFCP game with the centralized scheme. From the simulation results, we can see that the distributed solution and the centralized solution are almost the same when $\alpha$ is adjusted to the optimal working point.

%%%%%%%%%%%%%%%%%%%%%%%%%%%%%%%%%%%%%%%%%%%%%%%%%%%%%%%%%%%%%%%%%%%
\section{Conclusions}
%%%%%%%%%%%%%%%%%%%%%%%%%%%%%%%%%%%%%%%%%%%%%%%%%%%%%%%%%%%%%%%%%%%
In this paper, we have studied the CSI feedback rate control problem in a single-cell wireless data network, where a multiple-antenna BS communicates with a number of co-channel users through a MMSE precoder. Specifically, we proposed a non-cooperative feedback-rate control game without and with price. The price function is a linear function of the CSI feedback rate. The existence of the Nash equilibrium of such a game is proved. Simulation results are performed over FDMA and CSMA protocols in the feedback channel. It shows that the distributed NFCP game with the proposed utility results in improving the overall throughput of wireless data networks, and the simple distributed algorithm can provide comparative performance in comparison of the centralized one by properly varying the pricing parameter.

%%%%%%%%%%%%%%%%%%%%%%%%%%%%%%%%%%%%%%%%%%%%%%%%%%%%%%%%%%%%%%%%%%%%%%%%%%%%%%%%%%%%%%%%%%%%%%%%%%%
%%%%%%%%%%%%%%%%%%%%%%%%%%%%%%%%%%%%%%%%%%%%%%%%%%%%%%%%%%%%%%%%%%%%%%%%%%%%%%%%%%%%%%%%%%%%%%%%%%%
\appendices
\section{Parameter Derivations in (\ref{Eq:qua})}
%%%%%%%%%%%%%%%%%%%%%%%%%%%%%%%%%%%%%%%%%%%%%%%%%%%%%%%%%%%%%%%%%%%%%%%%%%%%%%%%%%%%%%%%%%%%%%%%%%
Thought channel quantization, $\mu$ can be simply expressed by the following linear function
%%--
\begin{equation}
    \mu=x+yD_k.
\label{Eq:mu}
\end{equation}
%%--
The real channel output $\mathbf{h}_{k}$ and its corresponding quantized channel $\overline{\mathbf{h}}_{k}$ in (\ref{Eq:qua}) satisfies the following linear extreme conditions:
%%--
\begin{itemize}
  \item When there is no quantization errors, i.e., $\nu=D_k=0$, we have
  %%--
\begin{equation}
    \mu^2=x^2=1;
\label{Eq:muvalue}
\end{equation}
%%--
  \item When the quantization is completely inaccurate, i.e., $D_k=1$ and $\mu=0$, we get
    %%--
\begin{equation}
    \mu^2=(x+y)^2=0.
\label{Eq:nuvalue}
\end{equation}
%%--
\end{itemize}
%%--
Combining (\ref{Eq:muvalue}) and (\ref{Eq:nuvalue}), we may easily get $\mu=1-D_k$. Recalling (\ref{Eq:hk}), from (\ref{Eq:qua}), we can also have
%%--
\begin{equation}
    \mathrm{Var}\left[\overline{\mathbf{h}}_{k}\right]=\mathrm{Var}\left[\mu\mathbf{h}_{k}+\nu\mathbf{n}_{q}\right]
                 =\mathrm{Var}\left[\mu\mathbf{h}_{k}\right] + \mathrm{Var}\left[\nu\mathbf{n}_{q}\right]\Rightarrow 1-D_k=\mu^2+\nu^2,
\label{Eq:varrea}
\end{equation}
%%--
and thus, we can finally obtain $\nu=\sqrt{D_k(1-D_k)}$.
%%%%%%%%%%%%%%%%%%%%%%%%%%%%%%%%%%%%%%%%%%%%%%%%%%%%%%%%%%%%%%%%%%%%%%%%%%%%%%%%%%%%%%%%%%%%%%%%%%

\pagebreak
%%%%--------------
%%%%--------------

%%%%--------------
%\clearpage

%%----------------------------------------------------
%%----------------------------------------------------

%%---------------------------------------------------
%---------------------------------------------------
\clearpage
%%---------------------------------------------------
%---------------------------------------------------

%%---------------------------------------------------
%---------------------------------------------------
\clearpage
%%---------------------------------------------------
%---------------------------------------------------

%%---------------------------------------------------
%---------------------------------------------------

\clearpage
%%---------------------------------------------------
%---------------------------------------------------

%%---------------------------------------------------
%---------------------------------------------------

\clearpage
%%---------------------------------------------------
%---------------------------------------------------

%%---------------------------------------------------
%---------------------------------------------------
\end{document}